\documentclass[numbers,preprint,review,10pt,times]{elsarticle}

\usepackage{amssymb}

\begin{document}

\begin{frontmatter}

\title{Electromediated formation of DNA complexes with cell membranes and
its consequences for gene delivery}

\author{Jean-Michel Escoffre \fnref{fn1}}
\address{Institut de Pharmacologie et de Biologie Structurale, CNRS, Universit\'e de Toulouse, UPS,  Toulouse, France}
\author{Thomas Portet \fnref{fn1}}
\address{Institut de Pharmacologie et de Biologie Structurale and Laboratoire de Physique Th\'eorique, CNRS, Universit\'e de Toulouse, UPS,  Toulouse, France}
\author{Cyril Favard \fnref{fn1}}
\address{Institut Fresnel, CNRS, Universit\'es Aix-Marseille, Marseille, France}
\author{Justin Teissi\'e}
\address{Institut de Pharmacologie et de Biologie Structurale, CNRS, Universit\'e de Toulouse, UPS,  Toulouse, France}
\author{David S. Dean}
\address{Laboratoire de Physique Th\'eorique, CNRS, Universit\'e de Toulouse, UPS, Toulouse, France }
\author{ Marie-Pierre Rols \corref{cor1}}
\ead{Marie-Pierre.Rols@ipbs.fr, Fax : +33 (0) 561 175 994, Tel :+33
(0) 561 175 811}
\address{Institut de Pharmacologie et de Biologie Structurale, CNRS, Universit\'e de Toulouse, UPS,  Toulouse, France}
\cortext[cor1]{Corresponding author} \fntext[fn1]{The first three
authors made an equal contribution to this paper}

\begin{abstract}
Electroporation is a  physical method to induce the uptake of
therapeutic drugs and DNA, by eukaryotic cells and tissues. The
phenomena behind electro-mediated membrane permeabilization to
plasmid DNA have been shown  to be significantly more complex than
those for small molecules. Small molecules cross the permeabilized
membrane by  diffusion whereas  plasmid DNA  first interacts with
the electropermeabilized part of the cell surface, forming localized
aggregates. The dynamics of this process is still poorly understood
because direct observations have been limited to  scales of the
order of seconds. Here, cells are electropermeabilized in the
presence of plasmid DNA and monitored with a temporal resolution of
2 ms. This allows us to show that during the first pulse
application, plasmid complexes, or aggregates, start to form at
distinct sites on the cell membrane. FRAP measurements show that the
positions of these sites are remarkably immobile during the
application of further pluses.  A theoretical model is proposed to
explain the appearance of distinct interaction sites, the
quantitative increase in DNA  and also their immobility  leading to
a tentative explanation for the success of electro-mediated gene
delivery.

\end{abstract}


\end{frontmatter}


\section{Introduction}
\label{intro} Electropermeabilization is the phenomena by which the
application of an electric field across a biological membrane
renders it permeable to the passage of small and even macro
molecules such as DNA \citep{neu1989,wea1995}. It is exploited in
the clinical context where the permeabilization of cells to a
therapeutic agent is required, for example in gene therapy and
chemotherapy \citep{bel93,dau2008,rol2006}. Despite the use of
electropermeabilization in medical science, many questions remain
open as to the underlying biophysicochemical phenomena which
underpin its success. This is especially the case for the
permeabilization of membranes to DNA where a number of interesting
biological, chemical and physical factors remain to be understood.
Given the size of the DNA, if the permeabilization is due to pores
-- or conducting defects--, as suggested by the standard theory of
electroporation \citep{pas1979,pow1986}, the pores must be
relatively large due to (i) the relatively large size of the DNA and
(ii) the large  charge of DNA  as dielectric exclusion must be
overcome \citep{par1969}. The cell membrane has a much more complex
organization than a model lipid bilayer. One expects that the
location of regions permeabilized to DNA will be determined not only
by the local electric field but also by the local membrane
composition. In cell membranes, the regions most susceptible to
permeabilization may be those containing certain types of
transmembrane proteins or at the boundary between differing types of
lipid domains \citep{ant1980}. Another key, physicochemical factor
is how the field not only modifies the permeability of the membrane,
but also how it influences the movement of the DNA by
electrophoresis. Large polyelectrolytes are strongly advected by the
electric field \citep{gab1999}. Experimental studies \citep{gol2002}
have demonstrated that the physicochemical phenomena involved in
electro-mediated membrane permeabilization to plasmid DNA are indeed
significantly more complex than those for small molecules. Small
molecules cross the permeabilized membrane directly mainly by
post-pulse diffusion, whereas  plasmid DNA  first interacts with the
electropermeabilized part of the cell surface resulting in the
formation of localized aggregates. However the dynamics of this
initial interaction is only  partially understood because the direct
observations of \citep{gol2002} were limited to time scales
exceeding several seconds. In this paper, we have analyzed the
interaction between the cell membrane and strongly charged
macromolecules (4.7 kbp plasmid DNA) upon the application of a
permeabilizing electric field, at a  temporal resolution of 2 ms.
This enables us to address the following unresolved, and key,
questions (i) What are the phenomena leading to formation of the
sites where the DNA aggregates? (ii) What is the underlying dynamics
behind their formation? (iii) To what extent are the sites localized
in space, and over what time scales?  and (iv) What is the
biochemical mechanism behind  this localization? We put forward a
model for the transport of charged molecules in the presence of the
electric field which provides an interpretation for most of our
experimental findings on DNA and whose validity is also tested via
additional experiments using the, much smaller, charged molecule
Propidium Iodide.

\section{Materials and methods}
\label{MM}
\subsection{Cells}
\label{cells}
 Chinese hamster ovary (CHO) cells were used. The WTT clone was selected for its ability to grow in suspension or plated on Petri dishes or on a microscope glass coverslip. Cells were grown as previously described \citep{rol1992}. For microscopy experiments, $10^5$ cells were put on a Lab-tek chamber 12 hours before electric pulse treatment with 1 ml of culture medium.
\subsection{DNA staining}
 A 4.7 kbp plasmid (pEGFP-C1, Clonetech, Palo Alto, CA) carrying the green fluorescent protein gene controlled by the CMV promoter was stained stoechiometrically with the DNA intercalating dye TOTO-1 (or alternatively POPO-3) (Molecular Probes, Eugene, OR ) \citep{rye1992}. The plasmid was stained with 2.3x10$^{-4}$ M dye at a DNA concentration of 1$\mu$g/$\mu$l for 60 minutes on ice. This concentration yields an average base pair to dye ratio of 5. Even if the labeling is not covalent, the equilibrium is dramatically in favor of the linked form. Plasmids were prepared from {\em E. Coli} transfected bacteria by using Maxiprep DNA purification system (Qiagen, Chatsworth, CA).
\subsection{ Electropermeabilization apparatus.}
Electropulsation was carried out with a CNRS cell electropulsator
(Jouan, St Herblain, France) which delivers square-wave electric
pulses. An oscilloscope (Enertec, St. Etienne, France) was used to
monitor the pulse shape. The electropulsation chamber was built
using two stainless-steel parallel rods (diameter 0.5 mm, length 10
mm, inter-electrode distance 5 mm) placed on a Lab-tek chamber. The
electrodes were connected to the voltage generator. A uniform
electric field was generated. The chamber was placed on the stage of
an inverted digitized videomicroscope (Leica DMIRB, Wetzlar,
Germany) or a confocal microscope (Zeiss, LSM 5 life, Germany).

\subsection{Electropermeabilization}
Permeabilization of cells was performed by application of millisecond
electric pulses required to transfer genes and to load
macromolecules into cells. Cell viability was preserved when using
millisecond pulse duration by decreasing the electric field intensity
\citep{wol1994,gol1998}. Penetration of PI (100 $\mu$M in a low
ionic strength pulsing buffer: 10 mM phosphate, 1 mM MgCl2, 250 mM
sucrose, pH 7.4) was used to monitor permeabilization. 10 pulses of
20 ms duration and 0.5 kV/cm amplitude were applied at a frequency
of 1 Hz at room temperature. For plated cells, the culture medium
was removed and replaced by the same buffer described above.

\subsection{Microscopy}
Cells were observed with a Leica 100$\times$, 1.3 numerical aperture
oil immersion objective. Images (and optical sections) were recorded
with the CELLscan System from Scanalytics (Billerica, MA) fitted
with a cooled CCD camera (Princeton Instrument, Trenton, NJ). This
digitizing set up allowed a quantitative localized analysis of the
fluorescence emission as described previously \citep{gol2002}. This
was done along the cell membrane. Images were taken at a 1 Hz
frequency.
For fast kinetics studies, a Zeiss LSM 5 Live confocal microscope was used. All measurements were performed at room temperature. Image sequences were acquired at a frequency of 500 fps. 

\subsection{FRAP experiments}
FRAP experiments were performed on a Zeiss LSM 510 confocal
microscope. The 488 nm line of the Ar+ laser was used for excitation
of TOTO-1. We used a sequential mode of acquisition with a
63$\times$, 1.4 numerical aperture water immersion lens. After 50
pre-bleach scans, a region of interest (ROI) with a radius $w = 1
\mu$m was bleached, and fluorescence recovery was sampled on 150
scans, {\em i.e.} 40 s \citep{cau2005}.

\subsection{Actin cytoskeleton destabilization}
To examine the role of actin in the phenomenon of DNA membrane
interaction, cells were incubated at 37$^o$ C for 1 hour with 1
$\mu$M Latrunculin A (Sigma) in culture medium. This protocol is
known to efficiently disrupt the actin cytoskeleton \citep{sun2007}.


\section{Experimental results}
\subsection{Electropulsation experiments}
To study membrane permeabilization by electric fields, and the
associated processes of molecular uptake, we used PI and TOTO-1 labelled
plasmid DNA. Membrane permeabilization was induced by applying 20 ms
electric pulses of intensity  0.5 kV/cm. This protocol is known to
induce both membrane permeabilization and DNA transfer, accompanied
by  an associated gene expression \citep{gol2002}. Membrane
permeabilization was observed at the single cell level at a rate of
500 frames per second. This image acquisition frequency made it
possible to monitor the entire process of molecular uptake, both
during and after pulse application, with a good spatial resolution
using confocal microscopy. Particular attention was paid to the
molecular uptake occurring after a single permeabilizing pulse and a
series of 10 pulses.

\subsection{Electropulsation experiments with PI}
As shown in Figs.~\ref{fig1}A and \ref{fig1}C, and in agreement with
previous studies \citep{gol2002}, membrane permeabilization, as
detected by the uptake of PI, was observed at sites of the cell
membrane facing the two electrodes. The influx into the cells of PI,
deduced from  the associated fluorescence intensity, suggests free
transport across the permeabilized regions of the membrane into the
cytoplasm (no trapping in the region near the membrane is observed).
The uptake of PI into cells started at the moment the pulse was
applied and the concentration of PI in the cell increased for up to
a minute (Fig.~\ref{fig1}E), showing that the permeabilized membrane
state, due to defect regions such as metastable pores,  persists for
some time after the field is cut. Indeed, the majority of molecules
which were taken up entered the cell after the pulses.   Analyzing
the molecular uptake image by image led to the observation that PI
uptake was both asymmetric and delayed: it was higher at the anode
facing side of the cell and its entrance at the cathode facing side
occurred after a 4-5 second delay. This asymmetry can be explained
by the fact that PI has a charge +2e in solution and thus the
electrophoretic force drives it toward the cathode. The post-pulse
diffusion of PI into the cytoplasm of the permeabilized cell  can be
seen in Movie 1 (published in supporting information).

\subsection{Electropulsation experiments with DNA}
 Electro-induced uptake by cells in presence of plasmid DNA showed results that differed considerably from those for cells pulsed in presence of PI. DNA was only observed to interact with the membrane for electric fields greater than a critical value (0.25 kV/cm) which  induces their permeabilization, i.e. leading to PI uptake. In the low field regime   DNA simply flowed around the cells towards the anode. For permeabilizing fields, plasmid DNA was not observed to enter the cell during pulse application or during the minute that followed.  However, DNA was  seen to accumulate at distinct sites on the cathode facing side of the permeabilized cell membrane (Fig.~\ref{fig1}D). At these sites the  DNA appeared to form aggregates, which became visible as soon as the electric pulse was triggered  (Fig.~\ref{fig2}A, \ref{fig2}B). Intriguingly, the number and apparent size of these sites did not increase with the number of pulses applied. During the application of a single pulse, the fluorescence of the sites increased linearly in time  showing that the quantity of DNA at each site increased linearly (Fig.~\ref{fig1}F), but no such fluorescence increase was observed  in the absence of the electric field. Therefore the accumulation of DNA at the interaction sites must be principally due to electrophoresis.

The presence of stable sites, where membrane plasmid DNA interaction
occurred, was observed across the entire cell population
(Fig.~\ref{fig1}D) (see Movie 2 in supporting information).  The
average distance between a site and its nearest neighbor was  found
to be about 1 $\mu$m. The diameter of the individual sites was found
to be in the range of 300-600 nm (lower range limit due to optical
diffraction).  As shown in Fig.~\ref{fig2}C, the total amount of DNA
in the membrane region also increased linearly with the number of
electric pulses. The final average amount of DNA localized in the
membrane region  correlated with the amount of DNA  reaching the
nucleus, as estimated by the rate of GFP expression,  between 2 and
24 hours after  pulse application (see supplementary information).

As shown in Figs.~\ref{fig1}E and \ref{fig1}F, part of the DNA
interacting with the membrane was observed to be desorbed after the
pulses but a large proportion stayed fixed, this could be due to an
interaction between the membrane and DNA but also because the DNA
forced into the pore electrophoretically is virtually immobile in
the cell in the absence of the electric field. This last observation
shows that the observed DNA membrane interaction could be, at least
partially, due to (i) electrophoretic accumulation along with (ii)
reduced mobility of the DNA in the pore region within the cell.

We also studied the mobility of the DNA complexes at the interaction
sites. In order to measure the in-membrane or lateral diffusion
constant of the complexes, we carried out Fluorescence Recovery
After Photobleaching (FRAP) experiments.  The FRAP results showed
that no exchange between DNA aggregates or with the bulk DNA took
place and that the  DNA was, on experimental time scales, totally
immobile, its lateral diffusion coefficient being less than
10$^{-16}$m$^2$s$^{-1}$.  This could be explained by the reduced
mobility of individual DNA molecules within the cell or because DNA
forms micro sized aggregates which are immobile due to their large
size. It is unlikely that this immobility is due to an intrinsic
immobility of pores, as the measured diffusion constant is orders of
magnitude smaller than the typical values reported in the literature
for  transmembrane objects, such as proteins.

One could argue that actin polymerization around inserted DNA may be
responsible for its immobilization at the membrane level as well as
in its subsequent traffic inside the cytoplasm. However, actin
cytoskeleton perturbation with Latrunculin A did not seem to modify
the features of DNA spot creation, their stability or immobility
(see supplementary information). This implies that, despite the fact
that actin causes a drastic decrease in the mobility of individual
large DNA molecules inside cells, the initial formation of DNA spots
and their immobility is not caused by the actin network.


\section{Theoretical analysis}
\subsection{Theoretical model}
In order to better understand the features of the electro-mediated
DNA uptake, we developed a theoretical model that we used to carry
out numerical simulations. We assumed that the cell membrane was an
infinitesimally thin non-conducting surface. This approximation is
valid where the membrane thickness is small compared to all other
length scales and the conductivities of the cell interior $\sigma_i$
and exterior electropulsation solution                   $\sigma_e$
are much greater than the conductivity $\sigma_m$ of the cell
membrane. An electric field of magnitude $|{{\bf E}}_0| = E_0$  is
applied in the $z$ direction, perpendicular to the electrodes, and
the local electric potential $\phi$ is given, in the steady state
regime, by the solution to Laplace's equation \citep{lan1975}
\begin{equation}
\nabla\cdot\sigma\nabla\phi = 0 \label{eqlap},
\end{equation}
with the boundary conditions $\phi=-E_0z$ as $|z|\to\infty$.  In
what follows we will assume that the DNA is advected by the field
given by the solution of Laplace's equation (\ref{eqlap}) and will
neglect additional fields that would be caused by local accumulation
of charge due to the DNA. This is because the applied fields are
relatively large and because we expect the DNA to be strongly
screened.  Once the electric field has been computed the DNA is
advected by the field but also diffuses. The local concentration of
DNA $c({\bf x},t)$ obeys the  electrodiffusion  equation
\begin{equation}
\frac{\partial c}{\partial t}=-\nabla\cdot{\bf j}\label{eqtran}
\end{equation}
where ${\bf j}$ is the thermodynamic current ${\bf j} = -D\nabla c
+\mu c {\bf E}$, with $D$ the local diffusion constant of the DNA,
which depends on its environment, and is denoted by $D_e$ outside
the cell, $D_m$ in the membrane and $D_i$ in the cell interior. The
term $\mu$ is the electrophoretic mobility and is defined via the
velocity ${\bf v}$ of the molecule in a (locally) uniform  field
${\bf E}=-\nabla\phi$ as
\begin{equation}
{\bf v} = \mu{\bf E}.\label{eqdefmu}
\end{equation}
As for the diffusion constant, the value of $\mu$ will also depend
on the local environment as it depends on electrolyte concentration
and viscosity. As the electrode system is in contact with a bath of
the transported molecule we impose the boundary conditions $\nabla
c=0$ as $c\to\infty$ on equation (\ref{eqtran}) (note that this is
compatible with the Ohmic behavior ${\bf j} =\mu c{\bf E}$ far from
the cell). We assume that $D_m$ and $\mu_m=0$, which means that DNA
cannot move into intact regions of membrane.

\subsection{Modeling  pores}
 We will assume that when the field is applied, the membrane becomes
permeabilized and micro sized pores are formed. For the ease of
computation we will investigate what happens when a circular pore is
formed at each face of the cell facing the electrodes, that facing
the anode in the angular region $\theta\in [0,\theta_p]$ and that
facing the cathode in the region $\theta\in [\pi-\theta_p,\pi]$. In
this region we assume that the conductivity is $\sigma_i$ and that
the diffusion constant is $D_i$, effectively we have taken away the
membrane from these regions and replace it by the cellular interior.
We numerically compute the evolution of the concentration of marker
molecules  as a function of time using the experimental applied
field protocols (we assume that the membrane charging process is
much quicker that any of the diffusion processes and thus change the
electric field instantaneously).

{\bf Estimation of theoretical parameters.} We were able to follow
the fluorescence of the DNA in the buffer solution in a uniform
electric field and thus measure its velocity. Using equation
(\ref{eqdefmu}). we estimated the electrophoretic mobility in
solution  to be   $\mu_e=10^{-8}m^2V^{-1}s^{-1}$. This agrees with
more precise measurements reported in the literature
\citep{ste1997}. The diffusion constant of DNA in aqueous solution
is estimated from the literature to be
$D_e=10^{-12}m^2s^{-1}$\citep{luk2000}.

The charge of the PI is taken to be   $q=+2e$ and its diffusion
constant in aqueous solution is estimated to be
$D_e=10^{-10}m^2s^{-1}$ (by estimating its effective radius and
using the Stokes formula for the diffusion constant of a solid
sphere). The electrophoretic mobility for PI is estimated via the
Stokes Einstein relation $\mu=\frac{qD}{k_BT}$ where  $k_B$ is
Boltzmann's constant and $T$  is the temperature.

In the cell interior we estimate that for PI the effective diffusion
constants and electrophoretic mobilities are smaller than their
external values by a factor of 10 but for the larger DNA molecules
they are smaller by a factor of 1 000  (in fact this factor is a
lower bound) due largely to the interaction between the DNA and the
actin network of the cell interior \citep{luk2000,dau2005}. Finally
we took the radius of the spherical vesicle to be $R=8\ \mu m$ as
estimated from the average cell size.

\subsection{Theoretical results and comparison with experiments}
\begin{figure}
\begin{center}
\centerline{\includegraphics[width=16cm]{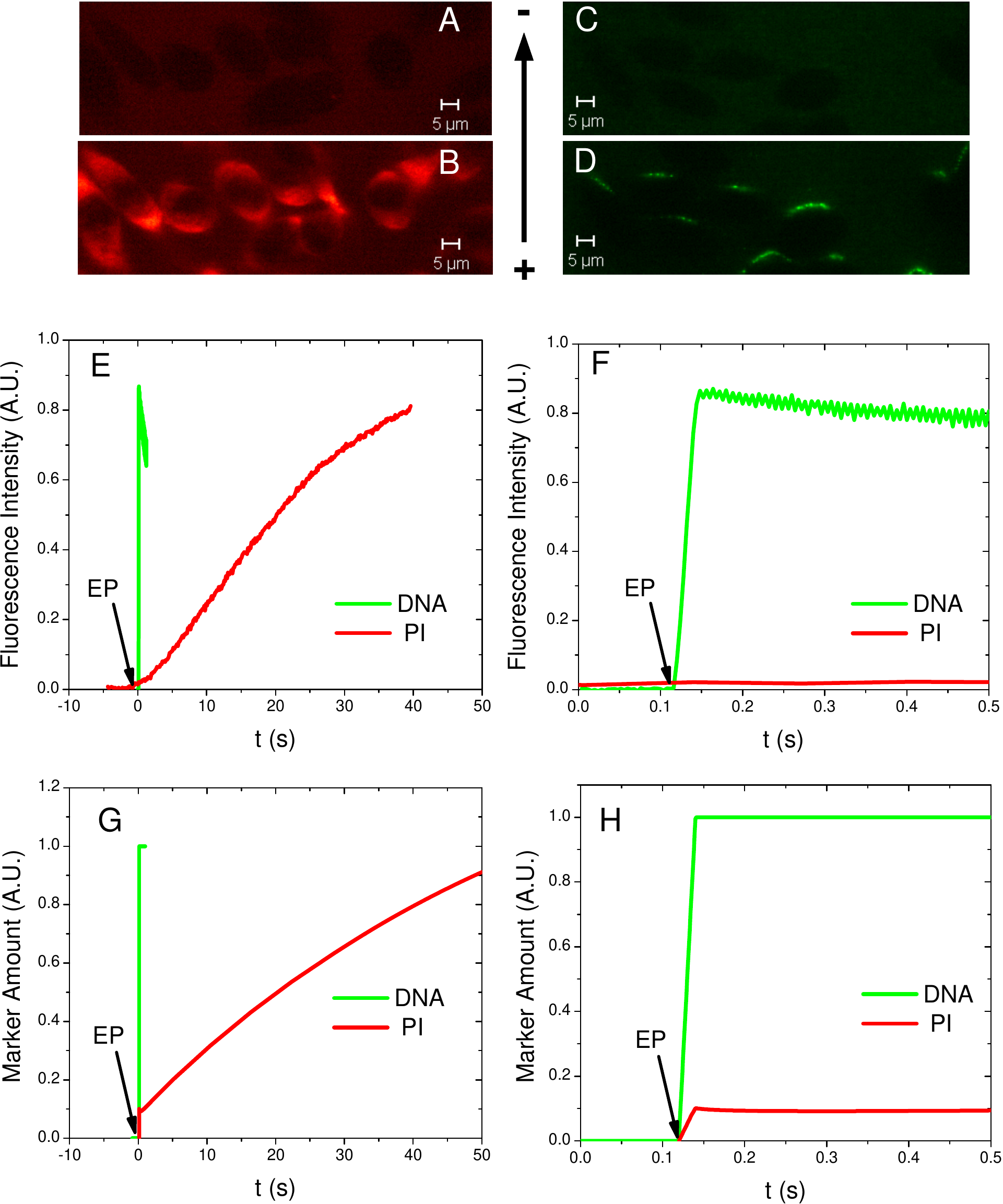}}
\caption{\textbf{Results and theoretical predictions of uptake of PI and DNA
by electropulsed cells.}  1A and 1B: cells in the presence of PI
and DNA before electropulsation, there is clearly no uptake of
either marker. 1C: uptake of PI by cells after being electropulsed.
1D: interaction of DNA and cells after being electropulsed. 1E: long
time behavior of PI (red) and DNA (green) uptake measured from the
corresponding fluorescence. 1F: same data shown over a shorter time
scale. 1G: long term uptake of PI and DNA as predicted by the
electrodiffusion model. 1H: corresponding behavior over shorter time
scale.}
\label{fig1}
\end{center}
\end{figure}

\begin{figure}
\begin{center}
\centerline{\includegraphics[width=16cm]{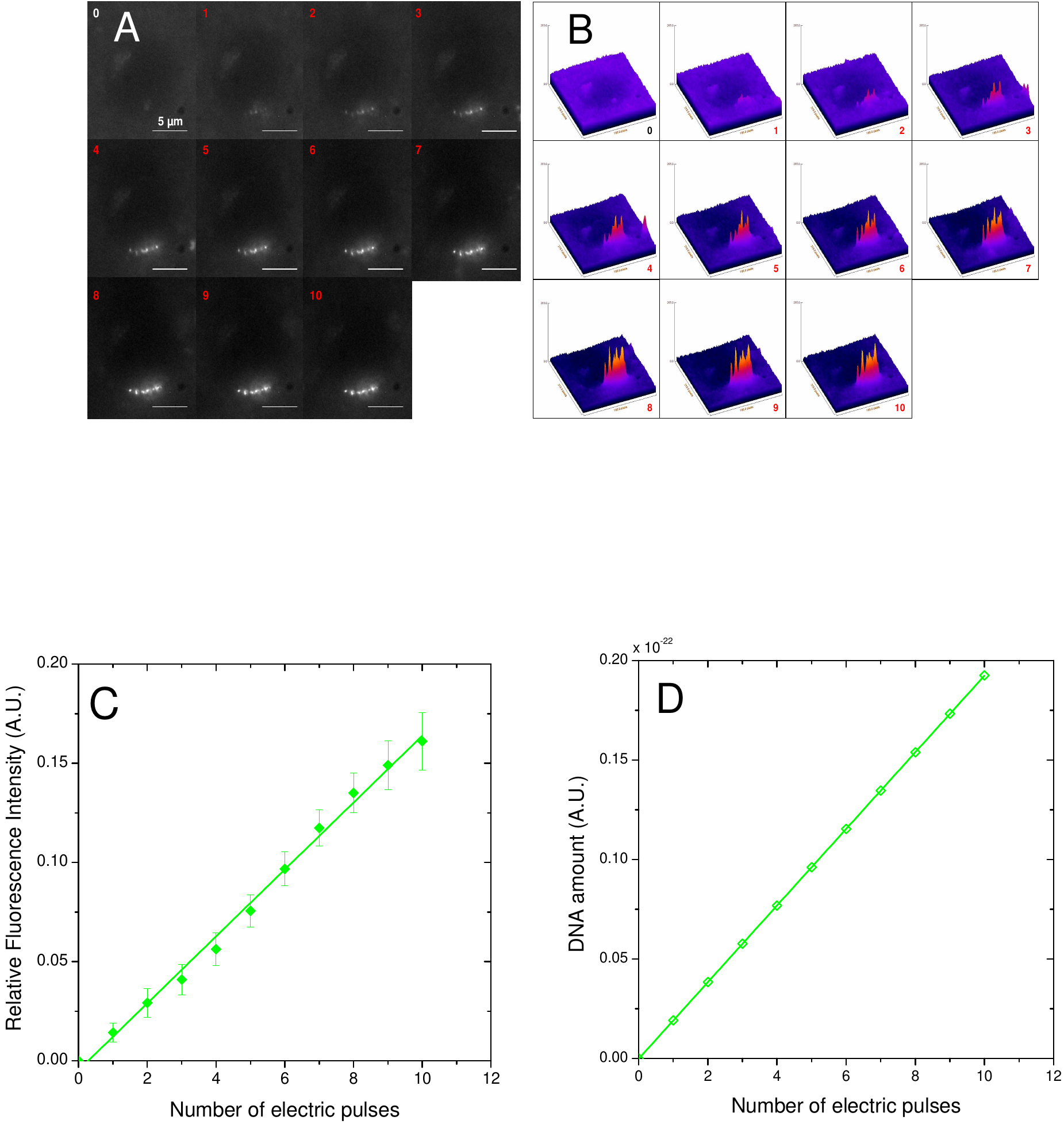}}
\caption{\textbf{Evolution of DNA fluorescence at interaction sites as a
function of the number of electropulses.} 
 2A raw imaging data.
2B: quantified increase in DNA fluorescence based on a digital
analysis of 2A. 2C: experimentally measured increase of total
fluorescence due to DNA uptake as a function of the number of
applied pulses. 2D: increase in DNA fluorescence as a function of
the number of pulses as predicted by the electrodiffusion model.}
\label{fig2}
\end{center}
\end{figure}

 We assume that the marker fluorescence is proportional to the concentration $c({\bf x},t)$. For the comparison with the experiments we took a pore angle $\theta_p=3.6^o$ as estimated by the size of the DNA/membrane interaction sites we observed.
We computed the  average fluorescence intensity inside the cell
using  $I_{cell} \propto c_{cell}$.  In Figs.~\ref{fig1}G and
\ref{fig1}H we show the predicted behavior for PI fluorescence for
one pulse with applied field $0.5\  kV\ cm{^{-1}}$  applied for $20\
ms$  and see that it  compares very well with that measured
experimentally (Fig.~1 E). Similarly we see  in the same figures the
comparison between theory and experiment is also good for  DNA (with
the same pulse protocol), with the  exception that the slight
decrease in DNA fluorescence after the pulse has been applied is
less pronounced in the theoretical curve of Fig.~\ref{fig1}F.
However some of the experimentally observed decrease may be due to
photo bleaching of the fluorescent marker. Note that although we
have only simulated one pore, if we assume that the pores behave
independently (as conduction channels in parallel) the overall form
of the increase in fluorescence for several pores should be
approximately of the same form, up to an overall change in the
fluorescence levels, as that for single pore.

In the case of PI, our model also reproduced the other qualitative
behavior seen in the experiments. Numerical calculations showed that
the PI enters through the anode facing side of the vesicle during
the pulse application and that its entry into the cathode facing
side is delayed, as expected from the physical arguments given in
the Results section.

For DNA, computations showed that just  after the field application
there is virtually no uptake of DNA at the anode facing side of the
cell, but that at the cathode facing side there is an accumulation
of DNA near the surface of the cell in the region where the pore is.
In our model the DNA accumulation, apparently, at the surface of the
cell, is due to the strong (as compared to the case of PI)
electrophoretic force which pushes the DNA into the hole opposite
the cathode and then the much reduced mobility of the DNA in this
region. For DNA most of the contribution to the total cell
fluorescence comes from the region close to the membrane surface,
thus justifying our comparison of Figs.~\ref{fig1}E, \ref{fig1}F
with Figs.~\ref{fig1}G, \ref{fig1}H.

\subsection{Constant number of interaction sites}
 The experimental results show that after the first interaction sites become visible,  no further sites appear during subsequent pulsation. An explanation for this
is that the pores  formed are conducting and thus the electric field
across the membrane is lowered in the non-conducting regions of the
membrane. A simple criterion for determining whether a membrane can
be locally permeabilized is if the stress caused by the electric
field causes the local surface tension to exceed the lysis tension
of the membrane. In a simple electric model for the membrane this
turns out to be equivalent to the local transmembrane voltage
$\Delta\phi$ exceeding a critical value $\Delta\phi_c$, which has
values typically between $250-1000\ mV$ \citep{tei93,por2009}. To
see how the presence of a pore reduces the transmembrane potential
elsewhere we consider the simplified case of an infinitesimally
thin, non-conducting, infinite flat membrane with a conducting pore
of radius $a$. The potential drop across the membrane at a distance
$r$ from the center of the  pore is \citep{win1987} $\Delta\phi(r) =
\frac{2\Delta\phi_o}{\pi}{\rm{arctan}}(\frac{\sqrt{r^2-a^2}}{a})$,
where $\Delta\phi_o$ is the potential drop far from the pore (or in
the absence of the pore) and we assume that
$\Delta\phi_o>\Delta\phi_c$ . If pores can only be formed in regions
where $\Delta\phi>\Delta\phi_c$  one sees that this critical
potential drop cannot be exceeded  within in a radius
$r_c=\frac{a}{\cos\left(\frac{\pi\Delta\phi_c}{2\Delta\phi_o}\right)}$
from the center of the pore. Although a very approximate estimate,
this shows that we should expect pores to be separated from each
other by a distance of the order of the pore size. This is indeed
the case as can be seen in Fig.~\ref{fig2}B, where the fluorescence
peaks widths are of the same order as the inter-peak distance.


\section{Discussion}
Despite the complexity of the system studied here, a number of our
experimental observations can be explained qualitatively and to an
extent quantitatively on a largely physical basis. According to the
standard theory of electropermeabilization, the effect of the
electric field is to render the cell membrane permeable to external
molecules via the formation of micro sized pores. Another effect of
the electric field on the DNA is an electrophoretic one, DNA is
pushed toward the cathode facing side of the cell and, as our
numerical calculations have shown, if a sufficiently large pore is
present the DNA can be forced through it. However once the DNA is
inside the cell it stays, on experimental time scales, very close to
the surface either due to the reduced electrophoretic mobility and
diffusion constant of individual DNA molecules caused by the actin
cytoskeleton or because it forms aggregates which are immobile due
to their large size. The number of DNA interactions sites remains
constant even on increasing the number of pulses, we have argued
that this is because these sites are conducting, the electric field
elsewhere in the cell membrane is reduced by the  presence of these
conducting sites and thus does not exceed the threshold value
necessary to form addition permeabilized regions and thus
interaction sites.

Results on cells where the actin cytoskeleton is disrupted also show
spot formation and so we conclude that the principal mechanism for
spot formation is the formation of aggregates where the DNA
molecules are bound together and thus diffuse as a macroscopic
object with a very small diffusion constant. A possible mechanism
for this aggregation is the presence of multivalent cations induced
by the high concentration of DNA in the pore region \citep{blo1999}.

To conclude we have been able to provide a  detailed explanation of
why gene therapy using electropulsation is successful. The process
of plasmid transfer through the cellular cytoplasm to the nuclear envelope
is a complex process \cite{de1,de2}. In principle
micro sized aggregates of DNA or vesicles filled with DNA could be
too large to pass through the pores formed by electroporation.
However individual DNA molecules, while they can pass through
electropores, have a limited mobility within the cell and may well
be totally degraded before reaching the nucleus. It is possible and
worth investigating the possibility that the actin cytoskeleton
reacts to the presence of DNA aggregates and plays an important role
in the subsequent intracellular transport. It seems reasonable that
only aggregates beyond a certain size (a few hundred nano-meters)
can induce a biological cellular response and can be transported by
the cell. In addition, the fact that the DNA is in aggregate form
means that the DNA in the center of the aggregate is relatively
protected from degradation. Therefore, for gene therapy purposes, it
is optimal for  DNA to  enter the cell as single molecules, but the
subsequent transport toward the nucleus is, for biological (possibly
by inducing a response of the actin cytoskeleton) and physical
(diminishing enzymatic degradation) reasons, optimized if the DNA is
in a micro-sized aggregate form. 

We thus see that a rather beautiful
and subtle, and to an extent fortuitous,  combination of biological,
chemical and physical factors may underpin the success of gene
therapy via electropulsation. As our understanding of these
underlying phenomena advances  we should be able to refine and
optimize the protocols used in  electro-mediated gene therapies.


\section{Acknowledgements}
Our group belongs to the CNRS consortium Celltiss. This work has
benefitted from the financial support of the Association Fran\c
caise contre les Myopathies, the Region Midi-Pyrenees and the
Institut Universitaire de France. We wish to thank Gabor Forgacs for
useful discussions on this work.

\section{Figures}

\clearpage

\end{document}